\documentclass[12pt, draftclsnofoot, onecolumn]{IEEEtran}
\usepackage{cite,graphicx,amsmath,amssymb}
\usepackage{subfigure}
\usepackage{citesort}
\usepackage{fancyhdr}
\usepackage{mdwmath}
\usepackage{mdwtab}
\usepackage{balance}
\usepackage{xcolor}
\usepackage{bm}

\usepackage{algorithm}
\usepackage{algorithmic}
\usepackage{multirow}
\usepackage{flafter}

\usepackage{multicol}
\usepackage{amsthm}
\usepackage{graphicx}
\usepackage{amsmath}
\usepackage{flafter}

\newtheorem{theorem}{Theorem}

\begin{document}


\title{Machine Learning for User Partitioning and Phase Shifters Design in RIS-Aided NOMA Networks}
\author{Zhong\ Yang,~\IEEEmembership{Student Member,~IEEE,}
Yuanwei\ Liu,~\IEEEmembership{Senior Member,~IEEE,}\\
Yue\ Chen,~\IEEEmembership{Senior Member,~IEEE}, and
Naofal\ Al-Dhahir,~\IEEEmembership{Fellow,~IEEE}

\thanks{ Part of this paper has been presented in IEEE Global Communication Conference (GLOBECOM) 2020~\cite{zhong2020globecom}.}
\thanks{ Z. Yang, Y. Liu and Y. Chen are with the School of Electronic Engineering and Computer Science, Queen Mary University of London, London E1 4NS, UK. (email:\{zhong.yang, yuanwei.liu, yue.chen\}@qmul.ac.uk)}
\thanks{ N. Al-Dhahir is with the Department of Electrical and Computer Engineering, University of Texas at Dallas, Richardson, TX 75080. (email:~aldhahir@utdallas.edu )}
}
\maketitle

\begin{abstract}
A novel reconfigurable intelligent surface (RIS) aided non-orthogonal multiple access (NOMA) downlink transmission framework is proposed. We formulate a long-term stochastic optimization problem that involves a joint optimization of NOMA user partitioning and RIS phase shifting, aiming at maximizing the sum data rate of the mobile users (MUs) in NOMA downlink networks. To solve the challenging joint optimization problem, we invoke a modified object migration automation (MOMA) algorithm to partition the users into equal-size clusters. To optimize the RIS phase shifting matrix, we propose a deep deterministic policy gradient (DDPG) algorithm to collaboratively control multiple reflecting elements (REs) of the RIS. Different from conventional training-then-testing processing, we consider a long-term self-adjusting learning model where the intelligent agent is capable of learning the optimal action for every given state through exploration and exploitation. Extensive numerical results demonstrate that: 1) The proposed RIS-aided NOMA downlink framework achieves enhanced sum data rate compared with the conventional orthogonal multiple access (OMA) framework. 2) The proposed DDPG algorithm is capable of learning a dynamic resource allocation policy in a long-term manner. 3) The performance of the proposed RIS-aided NOMA framework can be improved by increasing the granularity of the RIS phase shifts. The numerical results also show that reducing the granularity of the RIS phase shifts and increasing the number of REs are two efficient methods to improve the sum data rate of the MUs.
\end{abstract}

\section{Introduction}\label{section:introduction}

Due to the explosive increase of mobile devices, mobile data traffic has been growing dramatically in wireless networks. According to a white paper from Cisco~\cite{Cisco2019vni}, there will be 5.7 billion (71$\% $ of global population) mobile users by 2022, up from 5.0 billion in 2017, a compound annual growth rate (CAGR) of 2.8$\% $, which will generate 77.5 exabytes of mobile data traffic per month by 2022, up from 11.5 exabytes per month in 2017. In order to satisfy the higher requirements in data rates, lower latency, and massive connectivity in future generation wireless networks, promising technologies have been introduced and actively investigated, such as a reconfigurable intelligent surfaces (RIS), non-orthogonal multiple access (NOMA), and deep reinforcement learning (DRL).

An RIS reconfigures the wireless propagation environment via adjusting the passive beamforming adaptively. As an appealing complementary solution to enhance wireless transmissions, RIS are composed of massive low-cost and nearly passive reflective elements (REs) that can be flexibly deployed in the current wireless networks~\cite{Lianlin2017nature,Liaskos2018commag,Robin2019arxiv,Zhaorui2019arxiv,Yaoshen2019arxiv,Shimin2019arxiv}. NOMA is also a promising technology for massive user connectivity in future wireless communication networks. The integration of RIS with NOMA has already attracted significant attention both from academia and industry. RIS provides a new approach to enhance the NOMA performance by manipulating the wireless environment for mobile users who are blocked by obstacles, which motivates us to integrate RIS with NOMA downlink networks for further performance enhancement. Another promising recent technology is DRL, thanks to the rapid progression of fast and massively parallel graphical processing units (GPU). Reinforcement learning (RL) has been proven effective in ATARI games of Google DeepMind. The objective of RL algorithms is, for the defined intelligent agent, to intelligently take actions in an unknown environment, so as to maximize some notion of cumulative reward. Different from deep neural networks (DNN), which need large amounts of training data to model the complex environment, RL algorithms focus on finding the balance between exploration (of unknown environment) and exploitation (of known environment). In~\cite{Sun2019iot}, DRL is adopted for joint mode selection and resource management in green fog radio access networks. The authors in~\cite{Naparstek2019twc} proposed a deep multi-user RL based distributed dynamic spectrum access algorithm to maximize the formulated objective function. A DRL-based Online Offloading (DROO) framework is proposed in~\cite{Huang2019tmc} which implements a deep neural network as a scalable solution that learns the binary offloading decisions from experience. In this paper, we adopt DRL for RIS-aided NOMA downlink networks with the goal of answering the following key questions:

\begin{enumerate}

  \item[$\bullet$] $\bf Question~1:$ Do RIS-aided NOMA downlink networks significantly outperform RIS-aided orthogonal multiple access (OMA) downlink networks?

  \item[$\bullet$] $\bf Question~2:$ Does RIS dynamic phase shifting bring performance enhancement compared with a random phase shifting strategy?

  \item[$\bullet$] $\bf Question~3:$ Which parameter plays a more critical role in improving the performance of the RIS-aided NOMA downlink framework?

\end{enumerate}

\subsection{Motivations and Related Works}\label{subsection:motivationrelatedworks}

{\bf For the NOMA transmission aspect:} Different from conventional OMA techniques, such as orthogonal frequency division multiple access (OFDMA), which assigns subsets of subcarriers to an individual user, NOMA enables a base station (BS) to simultaneously transmit to several mobile users (MUs)~\cite{Yuanwei2017pieee,Vaezi2019tccn,Ding2017commag,Xiao2019jsac,Liu2016jcoml,Ding2016tvt,Wei2017tcom}. The key idea behind NOMA is to ensure that multiple users are served simultaneously within the same given time/frequency resource block (RB), utilizing superposition coding (SC) techniques at the transmitter and successive interference cancellation (SIC) at the receiver~\cite{Ding2017commag,Xiao2019jsac}. On one hand, for the NOMA scheme, user clustering has a significant impact on the tradeoff between the complexity of the SIC decoding and the performance. Recent research contributions have investigated user clustering for NOMA downlink for several aspects as the fairness guarantee~\cite{Liu2016jcoml}, the sum rate maximization~\cite{Ding2016tvt,Linsong2019arxiv}, and the transmit power minimization~\cite{Wei2017tcom}. On the other hand, for NOMA transmission, the channel gains for mobile users also affect the performance. If the mobile users are blocked, then we cannot apply the NOMA strategy.

{\bf For the RIS transmission aspect:} Previous research contributions on RIS mainly focus on sum transmit power minimization~\cite{Yang2020tcom}, sum-rate maximization~\cite{Huang2020jsac,Yu2020jsac}, spectrum efficiency~\cite{Xu2020tcom}, and energy efficiency~\cite{Zhaohui2020arxiv}. \cite{Yang2020tcom} investigate the sum transmit power minimization problem under signal-to-interference-plus-noise ratio (SINR) constraints of the users, which is solved by a dual method. A joint design of transmit beamforming and phase shifts is investigated in~\cite{Huang2020jsac} to maximize the sum rate of multiuser downlink multiple input single output (MISO) systems. The authors in~\cite{Yu2020jsac} implement the RIS by deploying programmable phase shifters to establish a favorable propagation environment for secure communication. In~\cite{Xu2020tcom}, the RIS is applied in a RIS-assisted full-duplex (FD) cognitive radio systems and joint beamforming design of the transmitter, the RIS and the receiver is optimized for maximizing the total spectral efficiency. A multiple RISs distribution is investigated in~\cite{Zhaohui2020arxiv} to maximize the energy efficiency dynamically. Different from~\cite{Yang2020tcom,Huang2020jsac,Yu2020jsac,Xu2020tcom,Zhaohui2020arxiv}, the authors in~\cite{Shen2019coml} maximize the secrecy rate for the RIS aided multi-antenna network. The performance of an RIS-aided large-scale antenna system is evaluated in~\cite{Han2019tvt} by formulating a tight upper bound on the ergodic spectral efficiency. Apart from the published papers cited above, there are more arXiv papers on RIS~\cite{Haibo2020arxiv,Yashuai2020arxiv}, intelligent reflecting surfaces (IRS)~\cite{Jianyue2019arxiv,Yuan2019arxiv}, and large reconfigurable intelligent surface (LRIS)~\cite{Keke2020arxiv}. In~\cite{Jianyue2019arxiv} and~\cite{Yuan2019arxiv}, the transmission power is minimized by optimizing the beamforming vector and the RIS phase shift matrix.

Due to RIS performance benefits, researchers also explored their integration with other key technologies such as millimeter wave communications~\cite{Peilan2019arxiv}, unmanned aerial vehicles (UAVs)~\cite{Yihan2020arxiv}, and OFDM~\cite{Beixiong2019arxiv}. The afore-mentioned potential benefits of RIS and NOMA motivate us to improve the system performance in RIS-aided NOMA downlink transmission~\cite{Zhiguo2019arxivirsnoma,Yiqing2019arxivirsnoma}. In~\cite{Zhiguo2019arxivirsnoma}, RIS-NOMA transmission is proposed to ensure that more users are served on each orthogonal spatial direction, compared to spatial division multiple access (SDMA). In~\cite{Yiqing2019arxivirsnoma}, the authors consider a multi-cluster MISO NOMA RIS-aided downlink communication network. In this paper, we propose RIS aided NOMA downlink networks as a promising solution for the above challenge. The application of machine algorithms to RIS brings promising advantages for improving the coverage and rate for NOMA aided downlink networks. Another advantage of NOMA aided downlink networks is enhancing the services for both line-of-sight (LOS: favorable signal propagation conditions) users and non-line-of-sight (NLOS: unfavorable signal propagation conditions) users, by smartly tuning the phase shifts, amplitude, and position.

{\bf For the RIS-aided NOMA aspect:} Recent research contributions have studied the potential benefits in RIS-aided NOMA networks~\cite{Xidong2019arxiv,Min2019arxiv}. In~\cite{Xidong2019arxiv}, the authors maximize the sum rate of mobile users by jointly optimizing the active beamforming at the BS and the passive beamforming at the RIS. The authors in~\cite{Min2019arxiv} minimize the downlink transmit power for a RIS-empowered NOMA network by jointly optimizing the transmit beamformers at the BS and the phase shift matrix at the RIS. The aforementioned RIS-aided NOMA contributions mainly focus on the static active beamforming at the BS and passive beamforming at RIS. However in practical scenarios, we often need to adjust the beamforming matrices in a long-term manner, which is complex in conventional optimization approaches.

{\bf For the DRL aspect:} Network intelligence is an extremely active area in the field of communication networks to meet the ever-increasing traffic demands. Thanks to the latest machine learning algorithms in artificial intelligence, especially RL algorithms and DRL algorithms, more and more RL algorithms are being adopted for communication network intelligence tasks~\cite{Yang2019iccdrl}, e.g., resource allocation, power control and demand prediction. In RIS-aided NOMA downlink networks, one critical challenge is to support ultra-fast wireless data aggregation, which pervades a wide range of applications in massive communication. RIS are envisioned as an innovative and promising technology to fufill the voluminous data transmission in future wireless networks, by enhancing both the spectral and energy efficiency~\cite{walid2019network,Liaskos2018commag,Shahu2018jsp}. RIS is a planar array composed of a large number of reconfigurable REs (e.g., low-cost printed dipoles), where each of the elements is able to induce certain phase shifting (by the attached intelligent controller) independently on the received signal, thus collaboratively adjusting the reflected signal propagation.

\subsection{Contributions and Organization}\label{subsection:contributionsorganization}

The beamforming strategy in RIS-aided NOMA networks still needs further investigation, especially when the network environment is complex and dynamic. For classical static optimization problems, conventional optimization approaches like branch and bound (BB)~\cite{Narendra1977Tcom}, alternating optimization (AO) techniques~\cite{zhang2020jsac} and Lagrange duality (LD) method~\cite{Xidong2020arxiv} can be used. However, considering a dynamic network where the beamforming matrix need to be calculated periodically makes the complexity of conventional approaches less practical. The above-cited references motivate us to further explore the potential benefits of applying DRL in RIS-aided NOMA downlink networks. In order to comprehensively analyze the network's performance enhancement brought by the RIS, we propose a RIS-aided multiple-input-single-output (MISO) network to study the sum rate of mobile uses, where some of the users are blocked, thus RIS is deployed efficiently to provide favorable wireless channels for them. Sparked by the aforementioned potential benefits of DRL, we explore the potential performance enhancement brought by DRL for RIS-aided NOMA downlink networks.

\begin{figure*} [t!]
 \centering
 \includegraphics[width=5in]{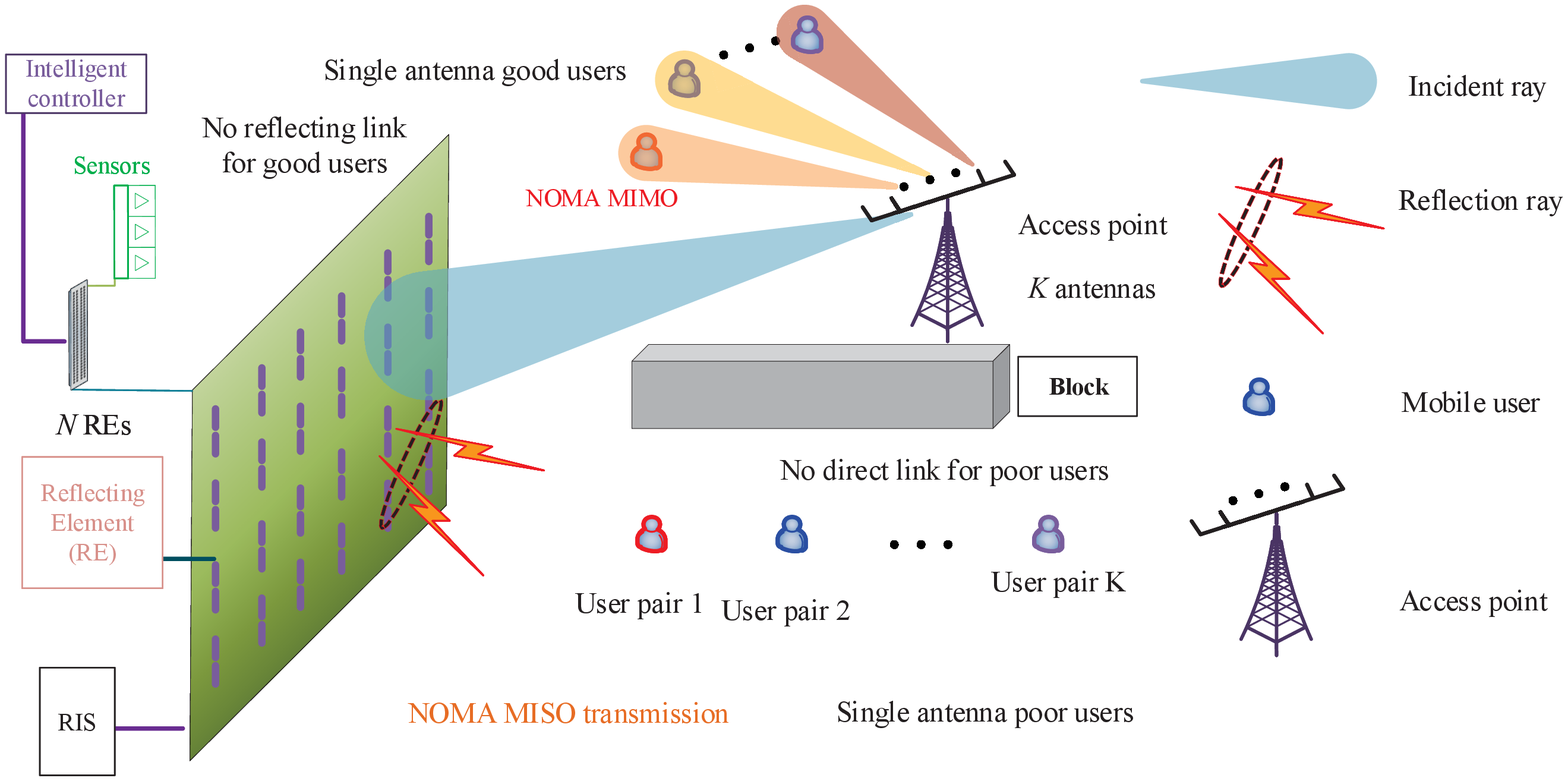}
 \centering
 \caption{An illustration of the RIS-aided NOMA downlink transmission.}\label{scenario}
\end{figure*}

In this paper, we consider a NOMA downlink network where half of the mobile users are blocked. The idea of applying RIS for smart communication is to improve the service of access points for both line-of-sight users and non-line-of-sight users~\cite{Chongwen2019arxiv}. The RL is adopted to find the long-term optimal setting of the REs of the RIS, i.e., their amplitudes and phases. As in Fig.~\ref{scenario}, a number of sensors collect environmental data. The sensors send their collected data to the controller through a wired link. In our contributions of the GLOBECOM paper~\cite{zhong2020globecom}, we invoke a DRL approach for the phase shifter design. In this paper, to better improve the performance, a novel modified object migration automation (MOMA) approach is utilized to partition the users into equal-size clusters. Our main contributions are summarized as follows

\begin{enumerate}
  \item We propose a RIS-aided NOMA downlink framework to study the performance enhancement of the RIS. We formulate a long-term sum rate maximization problem subject to the NOMA protocol, which is a combinatorial optimization problem.

  \item We apply a novel MOMA algorithm for rapid user clustering that is easy to implement and rapid to converge.

  \item We develop a deep deterministic policy gradient (DDPG) algorithm based solution for designing the phase shifts of the RIS. In the proposed DDPG based solution, we define the reward function using the sum rate of the mobile users, thus the formulated objective function finds the optimal trajectory of the intelligent agent.

  \item We demonstrate that the proposed RIS-aided NOMA downlink framework outperforms the conventional OMA framework in sum data rate. Increasing the number of REs is an efficient method to improve the performance of the proposed RIS-aided NOMA framework.
\end{enumerate}

The rest of this paper is organized as follows. In Section~\ref{section:systemmodel}, the system model for beamforming in the RIS is presented. In Section~\ref{section:omaup}, user partitioning is investigated. DDPG for phase shifts design is formulated in Section~\ref{section:ddpgresourceallocation}. Simulation results are presented in Section~\ref{section:numeralresult}, before we conclude this work in Section~\ref{section:conclusion}. Table~\ref{tableall} provides a summary of the notations used in this paper. Table~\ref{tableal_Acronyms} provides a summary of the acronyms in this paper.

\textit{Notation}: The expectation of a random variable is denoted as ${\mathbb E}\left(  \cdot  \right)$. The absolute value of a complex scalar is denoted as $\left|  \cdot  \right|$.

\begin{table*}[!t]
	\caption{LIST OF NOTATIONS}
	\centering
	\begin{tabular}{|c|l|c|l|}\hline Notation&Description&Notation&Description\\\hline
        $B$ & The bandwidth of the network & $K$ & The number of antennas \\\hline
        ${\bf{x}}\left( t \right)$ & The transmit signal of the AP & ${\bf{P}}\left( t \right)$ & The precoding matrix of the AP \\\hline
        $\widetilde {\bf{s}}$ & The transmit signal vector & ${\bf{h}}_{k,1}^{AU}$ & The channel vector \\\hline
        ${\bf{h}}_{RU}^{k,2} \in {{\mathbb C}^{1 \times N}}$ & The channel coefficients & ${n_{k,1}}$ & The addictive noise \\\hline
        ${\bf{Q}}\left( t \right)$ & The phase-shifting matrix of the RIS & ${\bf{h}}_{k,2}^{AR}\left( t \right)$ & The channel coefficients from the AP to the RIS \\\hline
        ${\rm{SIN}}{{\rm{R}}_{k,1}}\left( t \right)$ & The SINR of the good user & $\rho$ & The transmit power \\\hline
        ${\theta _n}\left( t \right)$ & The phase shifting of the $n$-th RE & $\bf{C}$ & User clustering \\\hline
        ${z_k}\left( t \right)$ & The position of the $k$-th user & ${{A_p}}$ & The query of a user \\\hline
        ${\Omega ^*}$ & The optimal user partition & ${A_1}$ & The reflecting amplitude of the RE \\\hline
        $s\left( t \right)$ & The state of the DDPG agent & $a\left( t \right)$ & The action of the DDPG agent \\\hline
        $lr$ & The learning rate the DDPG agent & ${V_\pi }\left( s \right)$ & The state-value function of the DDPG agent \\\hline
        ${Q^\pi }\left( {s,a} \right)$ & The action-value function of the DDPG agent & ${\pi _\theta }\left( {a|s} \right)$ & The action selection policy \\\hline
        $J\left( {{\pi _\theta }} \right){\rm{ }}$ & The expected reward & ${A^\pi }\left( {s,a} \right)$ & The advantage function \\\hline
        ${{{\rm P}_t}}$ & The transmit power of the AP & $\gamma$ & The discount factor \\\hline
        $H$ & The minibatch size & $U$ & The replay memory \\\hline
        ${{\bf{L}}_i}\left( {{\theta _i}} \right)$ & The loss fucntion & $O$ & The hidden layer size \\\hline
      \end{tabular}
	\label{tableall}
\end{table*}

\begin{table*}[!t]
	\caption{LIST OF KEY ACRONYMS}
	\centering
	\begin{tabular}{|l|l|l|l|}\hline
		Acronyms & Original text & Acronyms & Original text\\\hline
        NOMA & Mon-orthogonal multiple access & SC & Superposition coding \\\hline
        MU & Mobile users & SIC & Successive interference cancellation  \\\hline
        MOMA & Modified object migration automation & AP & Access point \\\hline
        DDPG & Deep deterministic policy gradient & IRS & Intelligent reflecting surface  \\\hline
        RIS & Reconfigurable intelligent surfaces & LRIS & Large reconfigurable intelligent surface \\\hline
        DRL & Deep reinforcement learning & LOS & Line-of-sight\\\hline
        RL & Reinforcement learning & NLOS & Non-line-of-sight \\\hline
        DNN & Deep neural networks & RE & Reflecting element\\\hline
        DROO & DRL-based Online Offloading & BB & Branch and bound \\\hline
        OMA & orthogonal multiple access & AO & Alternating optimization\\\hline
        OFDMA & Orthogonal frequency division multiple access & LD & Lagrange duality\\\hline
        BS & Base station & MISO & Multiple-input-single-output\\\hline
        RB & Resource block & CSI & Channel state information\\\hline
      \end{tabular}
	\label{tableal_Acronyms}
\end{table*}

\section{System Model}\label{section:systemmodel}

\subsection{Network Model}\label{subsection:networkmodel}

We consider a RIS aided NOMA downlink network operating over a bandwidth of $B$ Hz. The network architecture is illustrated in Fig.~\ref{scenario}, including an access point (AP) with $K$ antennas denoted by $\mathcal{K} = \{ 1,2, \cdots ,K\} $. There are $N$ REs on the RIS, where each RE is capable of independently reflecting the incident signal according to the channel state information (CSI), by controlling the amplitude and/or phase and thereby collaboratively achieve directional signal enhancement or nulling. We use $n \in {\mathcal N} = \left\{ {1,2, \cdots ,N} \right\}$ to index the REs of the meta-surfaces. As is illustrated in Fig.~\ref{scenario}, there are $K$ users close to the AP (these users are considered as good users). In addition, there are $K$ users far from the AP and are also blocked (these users are considered as poor users). To connect these poor users, we place an RIS near them to reflect the signal from the AP. To implement NOMA downlink transmission, we select one user from each group to form $K$ user clusters. The NOMA downlink strategy is adopted for each user cluster. There are obstacles between the users and the RIS, so there are no line-of-sight transmissions between the AP and users. As illustrated in Fig.~\ref{scenario}, the received signal from the RIS passes through three stages:

\begin{enumerate}

  \item[$\bullet$] ${\bf AP-RIS~transmission:}$ the RIS receives the signal from the AP. This is a multiple-input multiple-output (MIMO) NOMA downlink transmission.

  \item[$\bullet$] ${\bf RIS~reflection:}$ The RIS reflecting elements adjust the amplitude and/or phase of the received signal.

  \item[$\bullet$] ${\bf RIS-user~transmission:}$ The adjusted signal is transmitted to the users. This is a MISO NOMA downlink transmission

\end{enumerate}

\subsection{Wireless Channel Model}\label{subsection:signalmodel}


In the RIS-aided MISO NOMA downlink transmission, the transmitted signal of the AP is

\begin{equation}\label{transmittedsignal}
{\bf{x}}\left( t \right) = {\bf P}\left( t \right)\widetilde {\bf s},
\end{equation}
where ${\bf{P}} \in {{\mathbb C}^{K \times K}}$ is the precoding matrix generated by zero-forcing. Assuming that the channel matrix between the user and the AP is ${\bf{H}}$, and ${\bf H}^H$ represent the conjugate transpose of the channel matrix, then the precoding matrix $\bf{P = \left[ {{p_1}, \cdots {p_K}} \right]}$ is given by ${\bf{P}} = {\bf{H}}{\left( {{{\bf{H}}^H}{\bf{H}}} \right)^{ - {\bf{1}}}}$. ${\widetilde {\bf s}} \in {{\mathbb C}^{K \times 1}}$ denotes the signal vector given as follows:

\begin{equation}\label{transmittedsignal1}
{\widetilde {\bf s}} = \left[ {\begin{array}{*{20}{c}}
{\sqrt {{\alpha _{1,1}}} {s_{1,1}} + \sqrt {{\alpha _{1,2}}} {s_{1,2}}}\\
 \vdots \\
{\sqrt {{\alpha _{K,1}}} {s_{K,1}} + \sqrt {{\alpha _{K,2}}} {s_{K,2}}}
\end{array}} \right] \buildrel \Delta \over = \left[ {\begin{array}{*{20}{c}}
{{{\bf {\widetilde s_1}}}}\\
 \vdots \\
{{{\bf {\widetilde s_K}}}}
\end{array}} \right],
\end{equation}
where ${{s_{k,1}}}$ and ${{\alpha _{k,1}}}$ are defined as the transmitted information and the power allocation coefficient of the good user in the $k$-th cluster, respectively.

The received signal of the good user in cluster $k$ is given by

\begin{equation}\label{signalofuserk}
{y_{k,1}}\left( t \right) = {\bf{h}}_{k,1}^{AU}\left( t \right){\bf P}\widetilde {\bf s} + {n_{k,1}},
\end{equation}
where ${\bf{h}}_{k,1}^{AU} \in {{\mathbb C}^{1 \times K}}$ denotes the channel vector between the AP and the good user and ${n_{k,1}}$ represents the additive noise. We assume a Rayleigh fading channel model between the AP and the good user.

The received complex-valued signal at the poor user is given by

\begin{equation}\label{receivedsignal}
{y_{k,2}}\left( t \right) = {\bf{h}}_{RU}^{k,2}\left( t \right){\bf{Q}}\left( t \right){\bf{h}}_{k,2}^{AR}\left( t \right){\bf P}\left( t \right)\widetilde {\bf s} + {n_{k,2}},
\end{equation}
where ${\bf{h}}_{RU}^{k,2} \in {{\mathbb C}^{1 \times N}}$ represents the channel coefficients vector from the RIS to the poor user and ${n_{k,2}}$ represents the additive noise. ${\bf{Q}}\left( {{t}} \right) = diag\left[ {{q_1}\left( t \right),{q_2}\left( t \right), \cdots ,{q_N}\left( t \right)} \right]$ is the phase-shifting matrix of the reflection elements in the RIS, where ${q_n}\left( t \right) = \beta {e^{j{\theta _n}\left( t \right)}}$. ${\bf{h}}_{k,2}^{AR} \in {{\mathbb C}^{N \times M}}$ is the channel coefficients matrix from the AP to the RIS and ${n_{k,2}} \sim CN\left( {0,\sigma _{k,2}^2} \right)$ represents the additive Gaussian noise at the poor user.

As illustrated in Fig.~\ref{scenario}, the user from the far area is set as the weaker user because it only receives the signal from the RIS. Denote the $k$-th column of ${\bf P}$ by ${\bf p}_k$. Then, for the good user, the signal-to-interference-plus-noise ratio (SINR) is given by

\begin{equation}\label{SINRnearuser}
{\rm{SIN}}{{\rm{R}}_{k,1}}\left( t \right) = \frac{{\rho {{\left| {{\bf{h}}_{k,1}^{AK}\left( t \right){{\bf{p}}_k}\left( t \right)} \right|}^2}{\alpha _{k,1}}}}{{\sum\limits_{i = 1,i \ne k}^K {\rho {{\left| {{\bf{h}}_{k,1}^{AK}\left( t \right){{\bf{p}}_i}\left( t \right)} \right|}^2}}  + \sigma _{k,1}^2}},
\end{equation}
where $\rho$ is the transmit power. For the poor user, the SINR is given by (\ref{SINRfaruser}).

\begin{figure*}[t!]
    \normalsize
    \begin{align}\label{SINRfaruser}
    {\rm{SIN}}{{\rm{R}}_{k,2}}\left( t \right) = \frac{{\rho {{\left| {{\bf{h}}_{RU}^{k,2}\left( t \right){\bf{Q}}\left( t \right){\bf{h}}_{k,2}^{AR}\left( t \right){{\bf{p}}_k}\left( t \right)} \right|}^2}{\alpha _{k,2}}}}{{\rho {{\left| {{\bf{h}}_{RU}^{k,2}\left( t \right){\bf{Q}}\left( t \right){\bf{h}}_{k,2}^{AR}\left( t \right){{\bf{p}}_k}\left( t \right)} \right|}^2}{\alpha _{k,1}} + \sum\limits_{i = 1,i \ne k'}^K {\rho {{\left| {{\bf{h}}_{RU}^{k,2}\left( t \right){\bf{Q}}\left( t \right){\bf{h}}_{k,2}^{AR}\left( t \right){{\bf{p}}_i}\left( t \right)} \right|}^2}}  + \sigma _{k,2}^2}}.
    \end{align}
    \hrulefill \vspace*{0pt}
\end{figure*}

\subsection{Problem Formulation}\label{subsection:problemformulation}

In this paper, our objective is to maximize the sum rate of all the users in a long-term manner. There are two parameters that need to be optimized. In~\cite{Minchae2019ArXiv}, the authors assume that the phase shifter is capable of tuning the phase continuously ${\theta _n}\left( t \right) \in \left( {0,\pi } \right]$. However, in practical scenarios~\cite{Tiejun2014Science}, the phase can only be adjusted in a discrete manner, e.g.: $0,\pi/2,\pi,3\pi/2$. Therefore, in this paper, we assume that the phase of each RE can only be changed to $D-1$ phases. The formulated problem is formally given as:

\begin{subequations}\label{optimizationproblem1}
\begin{align}
\left( {{\bf{P1}}} \right)&\;\mathop {{\rm{max}}}\limits_{\bf{C,Q}} \sum\limits_{t = 1}^T {\sum\limits_{k = 1}^K {\sum\limits_{l = 1}^2 {{{\log }_2}\left( {1 + {\rm{SINR}}_{k,l}^{}\left( t \right)} \right)} } }  ,\label{objectivefunction1}\\
\mbox{s.t.} \quad
& C_1:\;c_k^m\left( t \right) \in \left\{ {0,1} \right\},\forall k \in \left[ {1,{K}} \right],m \in \left[ {1,M} \right],t \in \left[ {1,T} \right],\label{c1}\\
& C_2:\;{\theta _n}\left( t \right) \in \left\{ {\frac{1}{D}\pi , \cdots ,\frac{{D - 1}}{D}\pi } \right\},\forall n \in \left[ {1,N} \right],t \in \left[ {1,T} \right], \label{c2}\\
& C_3:\;{\rm{R}}_{j \to k}^m\left( t \right) \ge {\rm{R}}_{j \to j}^m\left( t \right),\pi _k^m\left( t \right) > \pi _j^m\left( t \right),\label{c3}\\
& C_4:\;\sum\limits_{k = 1}^K {\sum\limits_{l = 1}^2 {{\alpha _{k,l}}\left( t \right)} }  \le P .\label{c4}
\end{align}
where ${\bf{C}} = \left\{ {c_k^m\left( t \right),k \in {\cal K},m \in {\cal M},t \in {\cal T}} \right\}$, $\theta  = \left\{ {{\theta _n}\left( t \right),n \in {\cal N},t \in {\cal T}} \right\}$. Constraint (\ref{c3}) is to ensure that the SIC is performed successfully. Constraint (\ref{c4}) is the sum power constraint.
\end{subequations}

It is nontrivial to obtain the optimal solution of $\rm{\bf P1}$ because:

\begin{enumerate}

  \item[$\bullet$]  The objective function is non-concave with respect to each parameter.

  \item[$\bullet$]  The objective function is a long-term metric.

  \item[$\bullet$]  The constraint~(\ref{c3}) is non-convex.

\end{enumerate}

\begin{figure*} [t!]
 \centering
 \includegraphics[width=4.5in]{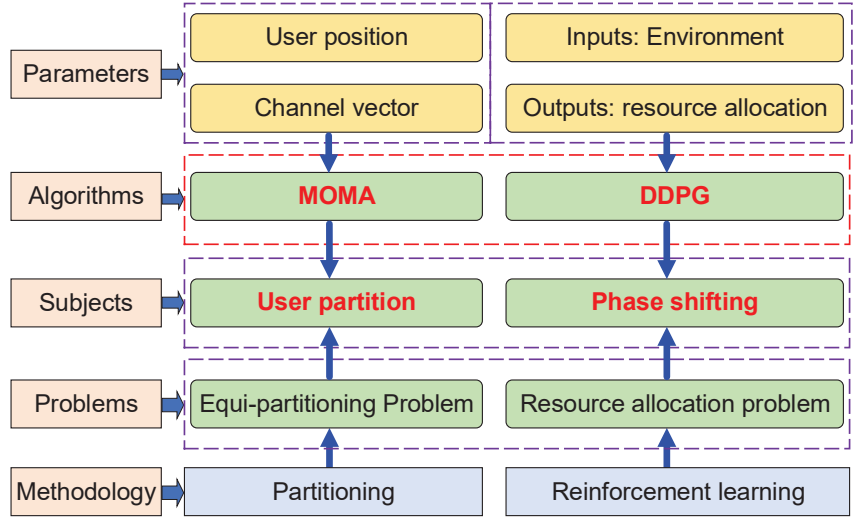}
 \centering
 \caption{The proposed MOMA and DDPG algorithm for user clustering and phase shifter design of the RIS. The users are partitioned into equal-size clusters by the proposed MOMA algorithm. The REs of the RIS are controlled by the intelligent agent of the DDPG algorithm, who is capable of learning the optimal phase shifts through exploration and exploitation.}\label{momaddpg}
\end{figure*}

The formulated maximization problem is a Markov decision process (MDP) problem. On the one hand, the channel power gain between the AP and user ${\bf{h}}_{k,1}^{AU}\left( t \right)$ is not known. On the other hand, the positions of the users follow a stochastic process, which makes it very challenging to solve $\rm{\bf P1}$. To solve {\bf (P1)}, we propose a MOMA algorithm for the user partitioning problem and a DDPG algorithm for the phase shifts design problem. The details of the structure of this paper are presented in Fig.~\ref{momaddpg}. RL is a machine learning algorithm that maximizes the long-term sum reward. Note that the objective function in $\rm{\bf P1}$ is also a long-term maximization problem. Therefore, we propose a RL based solution for $\rm{\bf P1}$ and we adopt the DDPG algorithm to design the phase shifts of the RIS.

\section{Modified Object Migration Automation for User Partitioning}\label{section:omaup}

In this section, we consider the fundamental problem of ``user partitioning" in NOMA transmission, which is a critical problem, because large numbers of mobile NOMA users cannot be processed in a single resource block (RB), but are better serviced when partitioned into orthogonal clusters. We partition the users into different clusters, where one beam is associated with each cluster, while the NOMA strategy is implemented in each cluster. We assume that for each beam, the bandwidth and power are the same, so we cluster the users into equal-size groups. Different from conventional object migration automation algorithm in~\cite{shirvani2019enhancing}, that the objects are all the same, in our scenario, there are two groups of users, one group of good users and another group of poor users. Therefore, we need to remove the users pairs if the two users comes from the same group, i.e., two good users or two poor users. Because the NOMA strategy is implemented for one good user and one poor user. To improve the user partitioning performance, we cluster the users according to both the positions of the users and the channel conditions. The position of the $k$-th user is denoted as $z_k\left( t \right) = \left( {x_k\left( t \right),y_k\left( t \right)} \right)$. As in {\bf Section~\ref{section:systemmodel}}, the channel of the $k$-th user is denoted as $g_k\left( t \right)$. Firstly, due to the fact that the user position and channel condition have different dimensions, we convert the expressions to be dimensionless using zero-mean normalization.


For the $K$ users, we denote their positions by $\left\{ {{z_1}\left( t \right),{z_2}\left( t \right), \cdots ,{z_K}\left( t \right)} \right\}$, where $z_k^{}\left( t \right) = \left( {x_k^{}\left( t \right),y_k^{}\left( t \right)} \right)$. We use zero-mean normalization to normalize the X and Y data as follows.

\begin{equation}\label{positionnormalization}
\begin{array}{l}
x{'_k}\left( t \right) = \frac{{{x_k}\left( t \right) - {u_x}\left( t \right)}}{{{\sigma _x}\left( t \right)}}\\
y{'_k}\left( t \right) = \frac{{{y_k}\left( t \right) - {u_y}\left( t \right)}}{{{\sigma _y}\left( t \right)}}
\end{array}
\end{equation}
where ${{u_x}}$ and ${{u_y}}$ denote the means of the X and Y data, respectively. The parameters ${{\sigma _x}}$ and ${{\sigma _y}}$ denote the standard deviations of the X and Y data, respectively.

For the channel gain $g_k^m\left( t \right)$ of user $k$, the normalized data is given as

\begin{equation}\label{channelnormalization}
g{'_k}\left( t \right) = \frac{{{g_k}\left( t \right) - {u_g}\left( t \right)}}{{{\sigma _g}\left( t \right)}},
\end{equation}
where ${{u_g}\left( t \right)}$ and ${{\sigma _g}\left( t \right)}$ denote the mean and standard deviation of the channel data.

After the normalization process, we obtain the data of user $k \in {\cal K}$ as ${d_k}\left( t \right) = \left[ {x{'_k}\left( t \right),y{'_k}\left( t \right),g{'_k}\left( t \right)} \right]$. The distance between users $k$ and $p$ is

\begin{equation}\label{distance}
\begin{array}{l}
dist\left( {k,p} \right) = \sqrt {{{\left| {{d_k}\left( t \right) - {d_p}\left( t \right)} \right|}^2}} \\
 = \sqrt {{{\left( {x{'_k}\left( t \right) - x{'_p}\left( t \right)} \right)}^2} + {{\left( {y{'_k}\left( t \right) - y{'_p}\left( t \right)} \right)}^2} + g{{\left( {x{'_k}\left( t \right) - g{'_p}\left( t \right)} \right)}^2}}.
\end{array}
\end{equation}


\begin{figure*} [t!]
 \centering
 \includegraphics[width=4.5in]{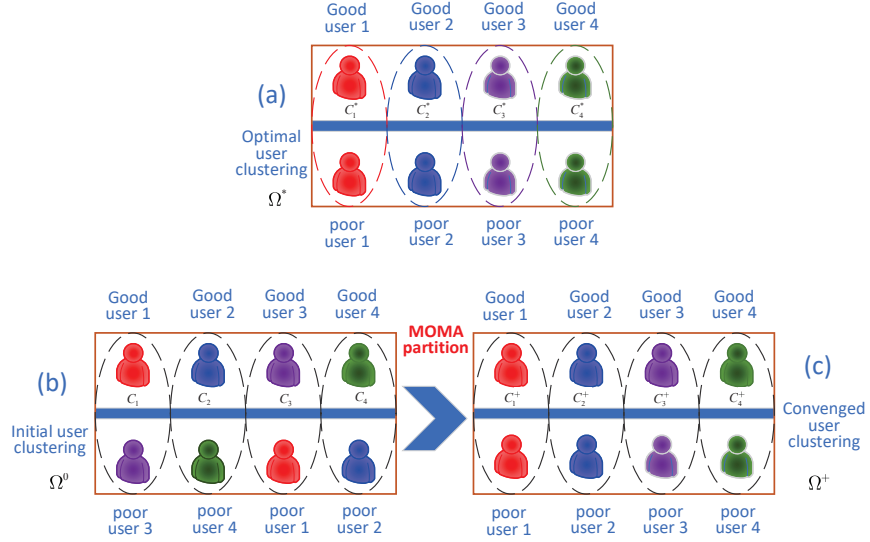}
 \centering
 \caption{The MOMA based user clustering (number of users is 8). The users are divided into two groups. (a) presents the optimal user clustering where good user 1 and poor user 1 constitute a NOMA group. (b) shows the initial user clustering which is not an optimal solution. (c) illustrates the converged user clustering of the proposed MOMA algorithm. }\label{omauc}
\end{figure*}

In the considered scenario, there are $2K$ users and the number of AP antennas is $K$. Therefore, we partition the users into $K$ groups. The optimal partition is denoted as ${\Omega ^*}$, where ${\Omega ^*} = \left\{ {G_1^*, \cdots ,G_K^*} \right\}$. The objective of the MOMA algorithm is to divide the $2K$ users into the $K$ groups by means of a partition ${\Omega }$ that converges to ${\Omega ^*}$ as the learning process proceeds. The elements of the proposed MOMA algorithm are given as:

\begin{itemize}
  \item Actions ($\mathcal{A}$): There are $K$ actions in the proposed MOMA $\left\{ {{\alpha _1}, \cdots ,{\alpha _K}} \right\}$.

  \item States ($\mathcal{S}$): For each action ${{\alpha _k}}$, there are $N$ states in the proposed MOMA denoted as $\left\{ {{\phi _{k1}}, \cdots ,{\phi _{kN}}} \right\}$. Therefore, there are $KN$ states in the proposed MOMA, represented as $\left\{ {{\phi _1}, \cdots ,{\phi _{KN}}} \right\}$. The automation states are separated into $K$ groups, each with a state depth of $N$. For groups $i,2, \cdots ,K$, when $i \ne \left\{ {1,K} \right\}$, the state index ranges from $\left( {i - 1} \right)N + 1$ through to $iN$.
  \item Reward (${\cal R}$): The reward is defined by the sum data rate in Eq.~(\ref{optimizationproblem1}). When an action is rewarded, the automation moves a step towards the innermost state of the action. However, if the automation is in states $1, \cdots ,\left( {i - 1} \right)N + 1, \cdots ,\left( {N - 1} \right)K + 1$, it remains there. On the other hand, on receiving a penalty, the automation moves a step towards the boundary states of the automation, and if the automation is in states $N, \cdots ,iN, \cdots ,KN$, it then moves to the boundary-state of the subsequent action (mod $K$).

  \end{itemize}

\subsection{Time and Space Complexity of the Proposed MOMA}\label{section:omaup1}

The time complexity of the proposed MOMA algorithm is equal to the number of queries, which is, in our case, equal to the number of users. Therefore, the time complexity of MOMA is $O(2K)$. The space complexity of the proposed MOMA algorithm contains two parts: the memory locations to store the states of the users and memory locations to store the sizes of the clusters. Thus, the space complexity of MOMA is $O\left( {2K + K} \right) = O\left( {3K} \right)$. We also calculate the time and space complexities of the K-means algorithm in Table~\ref{tablecomplexitymoma} as a benchmark. According to Table~\ref{tablecomplexitymoma}, the time and space complexities of the proposed MOMA is less than that of the K-means algorithm.
\begin{algorithm}[!t]
\caption{Modified object migration automation (MOMA) for User Clustering in NOMA Downlink Networks}
\label{Qlearningmec}
\begin{algorithmic}[1]
     \REQUIRE
     \STATE A set of users ${\cal N} = \left\{ {{n_1}, \cdots ,{n_N}} \right\}$.
     \STATE A stream of queries $\left\{ {\left\langle {{A_p},{A_q}} \right\rangle } \right\}$
     \STATE Initialization $\left\{ {{\delta _i}} \right\}$
\FOR{A sequence of $T$ queries}
\IF{$\delta _p$ div $N$ = $\delta _q$ div $N$}
    \IF{$\delta _p$ mod $N$ $\ne$ 1}
    \STATE ${\delta _p} = {\delta _p} - 1$
    \ENDIF
    \IF{$\delta _q$ mod $N$ $\ne$ 1 }
    \STATE ${\delta _q} = {\delta _q} - 1$
    \ENDIF
\ELSE
    \IF{${\delta _p}\bmod N \ne 0 \wedge {\delta _q}\bmod N \ne 0$}
    \STATE ${\delta _p} = {\delta _p} + 1$, ${\delta _q} = {\delta _q} + 1$
    \ELSE[${\delta _p}\bmod N \ne 0$]
    \STATE ${\delta _p} = {\delta _p} + 1$
    \ELSE[${\delta _q}\bmod N \ne 0$]
    \STATE ${\delta _q} = {\delta _q} + 1$
    \ELSE
    \STATE $temp={\delta _p}$, ${\delta _p} = {\delta _q}$, $l =$ index of an unaccessed user in group of $N_q$ closest to the boundary, ${\delta _l} = temp$
    \ENDIF
\ENDIF
    \IF{$\delta _p$ and $\delta _q$ in same group}
    \STATE Break.
    \ENDIF
\ENDFOR
     \ENSURE The $K$ clusters of users. The state ${\delta _i}$ of user $i$, where if ${\delta _i} \in \left[ {\left( {k - 1} \right)N + 1,kN} \right]$.
 \end{algorithmic}
\end{algorithm}

\begin{table*}[t!]
\caption{Complexity analysis of investigated algorithm.}
    \centering
	\begin{tabular}{|c|c|c|}\hline
		Investigated algorithms & MOMA algorithm& K-means algorithm\\\hline
        Time complexity & $O(2K)$ & $O(2{K^2})$ \\\hline
		Space complexity & $O(3K)$ & $O(4K)$  \\\hline
	\end{tabular}
	\label{tablecomplexitymoma}
\end{table*}

\section{DDPG Based RIS Phase Shifter Design}\label{section:ddpgresourceallocation}

After computing the user partitioning, we adopt RL to obtain a long-term phase shifts solution for the RIS. The ``State Space", ``Action Space", and ``Reward Function" for the proposed DDPG approach are given as follows:

\begin{enumerate}

  \item[$\bullet$] $\bf {State~Space:}$ The state space is defined by the phase shift matrix of the RIS. For a RIS with size ${N_{\rm RIS}} \times {N_{\rm RIS}}$, the state at time $t$ is defined as

        \begin{equation}\label{statedefinition}
        s\left( t \right) = \left[ {\begin{array}{*{20}{c}}
        {{A_1}{e^{j{\theta _1}}}}& \cdots &0\\
         \vdots & \ddots & \vdots \\
        0& \cdots &{{A_{{N_{\rm RIS}}}}{e^{j{\theta _{{N_{\rm RIS}}}}}}}
        \end{array}} \right],
        \end{equation}
        where ${A_1} =  \cdots  = {A_{{N_{\rm RIS}}}} = 1$ indicates that the receive signal is reflected by the RIS. ${\theta _1} =  \cdots  = {\theta _{{N_{\rm RIS}}}} \in \left\{ {\frac{1}{D}\pi , \cdots ,\frac{{D - 1}}{D}\pi } \right\}$ denotes the set of all possible discrete phase shift choices.

        \item[$\bullet$] $\bf {Action~Space:}$ The action matrix has the same dimension as the state matrix. The function of the action matrix is to change the diagonal elements in the state matrix. For example, the following action matrix at time $t$

        \begin{equation}\label{actiondefinition1}
        a\left( t \right) = \left[ {\begin{array}{*{20}{c}}
        1& \cdots &0\\
        \vdots & \ddots & \vdots \\
        0& \cdots &0
        \end{array}} \right]
        \end{equation}
        changes the first diagonal element in the state matrix by ${\theta _1} = {\theta _1} + \frac{1}{D}\pi $. On the other hand, the action matrix
        \begin{equation}\label{actiondefinition2}
        a\left( t \right) = \left[ {\begin{array}{*{20}{c}}
        -1& \cdots &0\\
        \vdots & \ddots & \vdots \\
        0& \cdots &0
        \end{array}} \right]
        \end{equation}
        changes the first diagonal element in the state matrix by ${\theta _1} = {\theta _1} - \frac{1}{D}\pi $.
        \item[$\bullet$] $\bf {Reward~Function:}$ The reward function is defined by the sum rate difference between the states; i.e.
\begin{equation}\label{rewarddefinition}
r\left( {s,a,s'} \right) = \left\{ {\begin{array}{*{20}{l}}
{{\rm{R}}\left( {s'} \right) - {\rm{R}}\left( s \right)}\\
0
\end{array}} \right.\begin{array}{*{20}{c}}
{if\;{\rm{R}}\left( {s'} \right) > {\rm{R}}\left( s \right)}\\
{{\rm{Others}}}
\end{array}.
\end{equation}
where ${\rm{R}}\left( s \right) = \sum\limits_{k = 1}^K {\sum\limits_{l = 1}^2 {{{\log }_2}\left( {1 + {\rm{SINR}}_{k,l}^{}\left( t \right)} \right)} } $ represents the data rate at state $s$. The state $s$ corresponds to one time slot $t$. $s'$ represents the next state after state $s$ when taking an action.

\item[$\bullet$] $\bf {Exploration~and~exploitation:}$ During the training phase, the action for each state is selected according to $\epsilon$-greedy, softmax and the Exponential-weight algorithm (Exp3 strategy)~\cite{Exp3exex}. The $\epsilon$-greedy based action selection is given by

    \begin{equation}\label{greedyactionselection}
    {a_n}\left( t \right) = \left\{ \begin{array}{l}
    \mathop {\arg \max }\limits_{\widetilde a \in \left\{ {0,1, \cdots ,K} \right\}} Q\left( {\widetilde a} \right)\\
    {\rm random~action}
    \end{array} \right.\begin{array}{*{20}{c}}
    {{\rm with~probability}\left( {1 - \varepsilon } \right)}\\
    {{\rm with~probability}~\varepsilon }
    \end{array}.
    \end{equation}

    The softmax based action selection is given by

    \begin{equation}\label{softmaxactionselection}
    \Pr \left( {{a_n}\left( t \right) = a} \right) = \frac{{{e^{{{Q\left( a \right)} \mathord{\left/
     {\vphantom {{Q\left( a \right)} \tau }} \right.
     \kern-\nulldelimiterspace} \tau }}}}}{{\sum\nolimits_{\widetilde a \in \left\{ {0,1, \cdots ,K} \right\}} {{e^{{{Q\left( {\widetilde a} \right)} \mathord{\left/
     {\vphantom {{Q\left( {\widetilde a} \right)} \tau }} \right.
     \kern-\nulldelimiterspace} \tau }}}} }},
    \end{equation}
    where $\tau$ is the ``temperature''. High temperatures cause the actions to be equiprobable. Low temperatures cause a greater difference in selection probability for actions that differ in their value estimates.

    The Exp3 based action selection is given by

    \begin{equation}\label{explorationandexploitation}
    \Pr \left( {{a_n}\left( t \right) = a} \right) = \frac{{\left( {1 - \alpha } \right){e^{\beta Q\left( a \right)}}}}{{\sum\nolimits_{\widetilde a \in \left\{ {0,1, \cdots ,K} \right\}} {{e^{\beta Q\left( {\widetilde a} \right)}}} }} + \frac{\alpha }{{K + 1}},
    \end{equation}
    where $K$ denotes the number of actions, while $\alpha$ and $\beta$ are the experience parameters. The Exp3 strategy for action selection balances between the softmax and $\epsilon$-greedy based action selection strategies.
\item[$\bullet$] $\bf {Dynamic~learning~rate:}$ Reducing the learning rate as the training progresses is an efficient method for enhancing the RL training progress. In this paper, we compare the following three learning rate decay strategies: iteration-based decay, step decay and exponential decay. For iteration-based decay, the learning rate decreases with the iterations according to the relation

    \begin{equation}\label{iterationbaseddecay}
    lr = \frac{{lr_0}}{{1 + kt}},
    \end{equation}
    where $lr_0$ is the initial learning rate, $k$ is a hyper-parameter and $t$ is the iteration number. For the step decay strategy, the learning rate decreases by a factor every few steps according to    \begin{equation}\label{stepdecay}
    lr = lr_0 * dro{p^{\frac{{step}}{{step\_drop}}}}.
    \end{equation}
    where $drop$ is the decay factor, $step$ is learning step of the agent, $step\_drop$ is the number of steps for each decay. For the exponential decay strategy, the learning rate is given by
    \begin{equation}\label{exponentialdecay}
    lr = {lr_0} * {e^{ - kt}}.
    \end{equation}

\end{enumerate}

In order to apply the RL algorithm for RIS-assisted networks, one key issue is to define the qualified reward function for the agent, which defines the reward and the cost after taking each action in the system. The reward and cost measure the quality of taking every action.
\begin{figure*} [t!]
 \centering
 \includegraphics[width=4.5in]{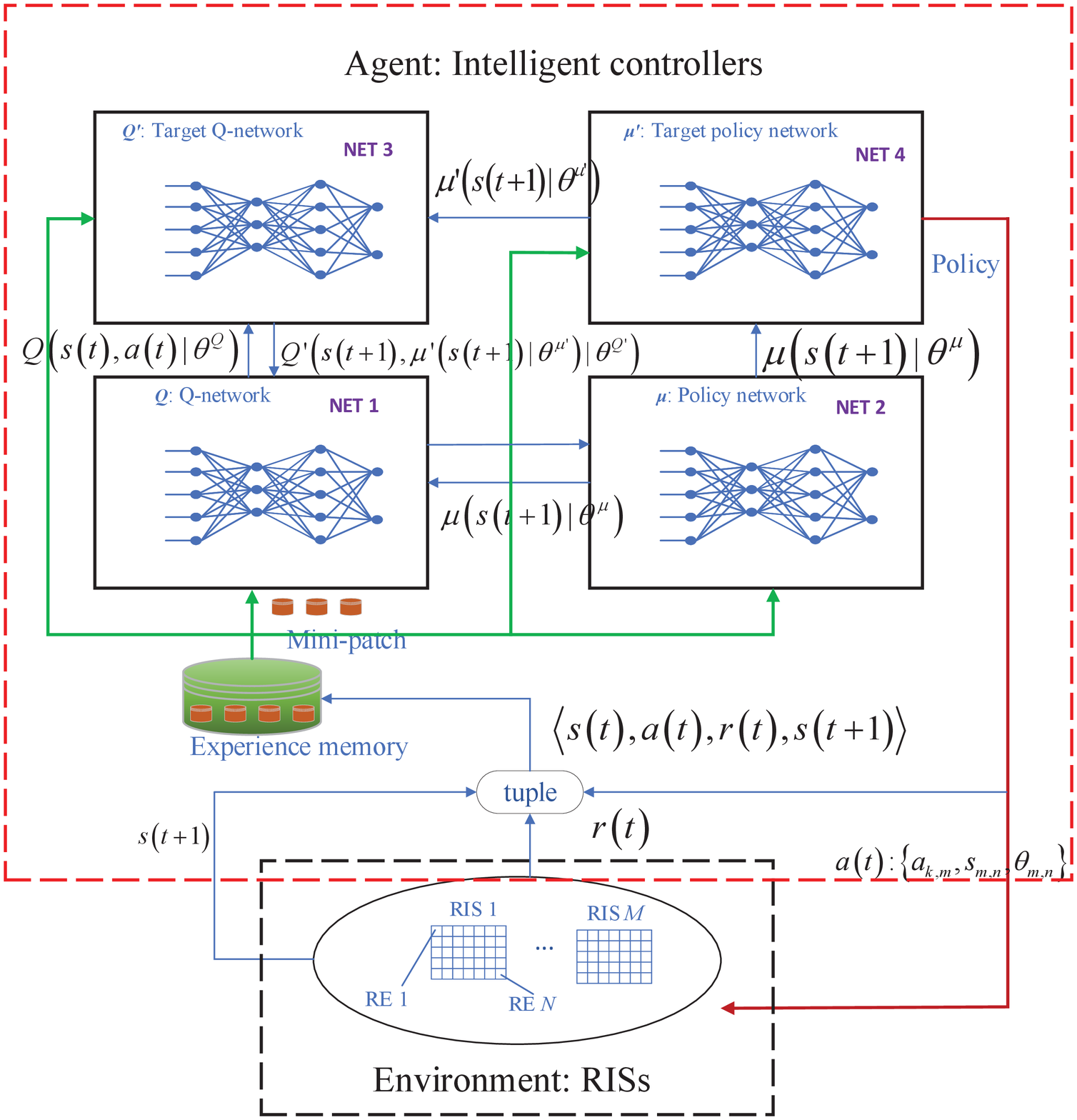}
 \centering
 \caption{The framework of the proposed DDPG algorithm for RIS-aided NOMA downlink transmission.}\label{DDPGRIS}
\end{figure*}

Based on the definition of the reward function, we reformulate the proposed problem as the following long-term expected reward maximization problem

\begin{equation}\label{reformulate}
\max ~ {V_\pi }\left( s \right) = {\mathbb E}\left[ {\sum\limits_{t = 0}^{ + \infty } {{\gamma ^t}r\left( t \right)|{s_0} = s} } \right].
\end{equation}
The parameter $\gamma  \in \left[ {0,1} \right]$ denotes the discount factor, where $\gamma  \to 0$ means that the short-term reward has more weight than the long-term reward, while $\gamma  \to 1$ means the opposite.

Value function approaches attempt to find a policy that maximizes the return by maintaining a set of estimates of expected returns for some policy. Given a state $s$, an action $a$ and a policy $\pi$, the action-value of the pair $\left( {s,a} \right)$ under $\pi$  is defined by

\begin{equation}\label{Qvaluepi}
{Q^\pi }\left( {s,a} \right) = {\mathbb E}\left( {r|s,a,\pi } \right).
\end{equation}

Policy gradient (PG) algorithms are widely used in RL algorithms with continuous action spaces. The basic idea is to represent the policy by a parametric probability ${\pi _\theta }\left( {a|s} \right) = {\mathbb P}\left( {a|s;\theta } \right)$ that stochastically selects action $a$ in state $s$ according to parameter vector $\theta$. The goal of PG is to derive a policy ${\pi _\theta }\left( {a|s} \right)$ and obtain the best $\theta$. We utilize the average reward per time-step to measure the quality of a policy ${\pi _\theta}$

\begin{equation}\label{weightvector1}
\begin{aligned}
J\left( {{\pi _\theta }} \right) &= \int\limits_S {{\rho ^\pi }\left( s \right)\int\limits_A {{\pi _\theta }\left( {a|s} \right)R\left( {a|s} \right)dads} }  \\
&  = {{\mathbb E}_{s \sim {\rho ^\pi },s \sim {\pi _\theta }}}\left[ {R\left( {a|s} \right)} \right].
\end{aligned}
\end{equation}
where ${{\mathbb E}_{s \sim {\rho ^\pi },s \sim {\pi _\theta }}}\left[ {R\left( {a|s} \right)} \right]$ represents the expected value with respect to the discount state distribution ${{\rho ^\pi }\left( s \right)}$.

\begin{figure*}[!t]
    \normalsize
    \begin{align}\label{gradientupdate}
    {\nabla _{{\theta _i}}}{{\bf L}_i}\left( {{\theta _i}} \right) = {{\mathbb E}_{s,a \sim \rho \left(  \cdot  \right);s' \sim \varepsilon }}\left[ {r + \gamma \mathop {\max }\limits_{a'} {Q^*}\left( {s',a';{\theta _{i - 1}}} \right) - Q\left( {s,a;{\theta _i}} \right){\nabla _{{\theta _i}}}Q\left( {s,a;{\theta _i}} \right)} \right].
    \end{align}
    \hrulefill \vspace*{0pt}
\end{figure*}

DDPG is an effective actor-critic, model-free RL algorithm for Markov decision process~\cite{Zhang2019twc} and obtains a long-term solution~\cite{Qiu2019iot}. As shown in Fig.~\ref{DDPGRIS}, there are four networks in the deep deterministic policy gradient (DDPG) structure, the function of each network is described below

\begin{enumerate}

  \item[$\bullet$] $\bf {Network 1: Q-network}$

        {\bf Input:} Mini-batches from the experience memory.

        {\bf Output:} $Q\left( {s\left( t \right),a\left( t \right)|{\theta ^Q}} \right)$.
The Q-value function satisfies the Bellman Equation

\begin{equation}\label{Bellmanequation}
{Q^*}\left( {s,a} \right) = {{\mathbb E}_{s' \sim \varepsilon }}\left[ {r + \gamma \mathop {\max }\limits_{a'} {Q^*}\left( {s',a'} \right)|s,a} \right].
\end{equation}

We use neural networks to approximate the optimal Q-value function. During the training part of the network, the loss function used is given as

\begin{equation}\label{lossfunction}
{{\bf L}_i}\left( {{\theta _i}} \right) = {{\mathbb E}_{s,a \sim \rho \left(  \cdot  \right)}}\left[ {{{\left( {{y_i} - Q\left( {s,a;{\theta _i}} \right)} \right)}^2}} \right],
\end{equation}
where $y_i$ is the Bellman equation given as

\begin{equation}\label{yi}
{y_i} = {{\mathbb E}_{s' \sim \varepsilon }}\left[ {r + \gamma \mathop {\max }\limits_{a'} {Q^*}\left( {s',a';{\theta _{i - 1}}} \right)|s,a} \right].
\end{equation}

After calculating the loss function, the gradient update with respect to the parameter $\theta$ is given in Eq.~(\ref{gradientupdate}).

    \item[$\bullet$] $\bf {Network~2: Deterministic~policy~network}$

        {\bf Input:} Q values from the Q-network.

        {\bf Output:} $\mu \left( {s\left( t \right)|{\theta ^\mu }} \right)$.

In addition, we adopt the difference between the Q-value function and the value function as the baseline to reduce the variance of the expected reward. The advantage function is given as follows

\begin{equation}\label{advantagefunction}
{A^\pi }\left( {s,a} \right) = {Q^\pi }\left( {s,a} \right) - {V^\pi }\left( s \right).
\end{equation}

Here, the gradient of Eq.~(\ref{weightvector1}) is given as

\begin{equation}\label{Jtheta}
{\nabla _\theta }J\left( \theta  \right) \approx \sum\nolimits_{t \ge 0} {\left[ {{Q^\pi }\left( {s,a} \right) - {V^\pi }\left( s \right)} \right]{\nabla _\theta }\log {\pi _\theta }\left( {{a_t}|{s_t}} \right)}.
\end{equation}

        \item[$\bullet$] $\bf {Network~3: Target~Q-network}$

        {\bf Input:} $Q\left( {s\left( t \right),a\left( t \right)|{\theta ^Q}} \right)$ and ${\mu '\left( {s\left( {t + 1} \right)|{\theta ^{\mu '}}} \right)}$.

        {\bf Output:} $Q'\left( {s\left( {t + 1} \right),\mu '\left( {s\left( {t + 1} \right)|{\theta ^{\mu '}}} \right)|{\theta ^{Q'}}} \right)$.

        In the target deterministic policy network, the agent chooses the action from the output of the network. The parameters are updated by ${\theta ^{Q'}} \leftarrow \tau {\theta ^Q} + \left( {1 - \tau } \right){\theta ^{Q'}}$.
        \item[$\bullet$] $\bf {Network~4: Target~deterministic~policy~network}$ The parameters are updated by ${\theta ^{\mu '}} \leftarrow \tau {\theta ^\mu } + \left( {1 - \tau } \right){\theta ^{\mu '}}$.
        {\bf Input:} $\mu \left( {s\left( t \right)|{\theta ^\mu }} \right)$

        {\bf Output:} ${\mu '\left( {s\left( {t + 1} \right)|{\theta ^{\mu '}}} \right)}$ and the action selection policy.

\end{enumerate}

The framework of DDPG for RIS-assisted communication is shown in Fig.~\ref{DDPGRIS}. The agent, (i.e., the central intelligent controller) receives the reward $r\left( t \right)$ and new state $s\left( t+1 \right)$ from the environment after taking an action $a\left( t \right)$ in state $s\left( t \right)$. Then, the transition tuple $\left\langle {s\left( t \right),a\left( t \right),r\left( t \right),s\left( {t + 1} \right)} \right\rangle $ is stored in the experience memory. A mini-batch contains $L$ transitions that are sampled from the experience memory to train the Q-network using the chain role as described above.

\begin{algorithm}[!t]
 \caption{DDPG Algorithm for Phase Shifters Design in RIS-aided NOMA Downlink}
 \label{Qlearningmec2}
 \begin{algorithmic}[1]
     \REQUIRE
     \STATE The number of time slots $T$ in one episode, the mini-batch size $H$, the learning rate $\tau$, the replay memory $U$, the hidden layer size $O$.
     \STATE Randomly initialize the switch and phase of the REs in the RIS, associate the users with the RIS randomly.
     \STATE Randomly initialize parameters ${{\theta ^Q}}$ and ${{\theta ^\mu }}$ in the Q-network $Q\left( {s\left( t \right),a\left( t \right)|{\theta ^Q}} \right)$ and deterministic policy network $\mu \left( {s\left( t \right)|{\theta ^\mu }} \right)$. Then, initialize the parameters ${{\theta ^{Q'}}}$ and ${{\theta ^{\mu '}}}$ in the target Q-network $Q'\left( {s\left( {t + 1} \right),\mu '\left( {s\left( {t + 1} \right)|{\theta ^{\mu '}}} \right)|{\theta ^{Q'}}} \right)$ and target deterministic policy network ${\mu '\left( {s\left( {t + 1} \right)|{\theta ^{\mu '}}} \right)}$.

     \FOR{each episode}{

         \STATE Generate an initial action sample from a uniform random distribution over all actions.
         \STATE Execute the sampled action from the last step and obtain the initial state.

     \FOR{each time slot}
     {
         \STATE Select the actions $a\left( t \right) = \mu '\left( {s\left( t \right)|{\theta ^{\mu '}}} \right) + {\mathcal N}\left( t \right)$ according to the current policy and exploration noise. \\

     \STATE Execute selected action $a\left( t \right):\left\{ {{a_{k,m}}\left( t \right),{s_{m,n}}\left( t \right),{\theta _{m,n}}\left( t \right)} \right\}$ and observe the reward $r\left( t \right)$ and new state $s\left( {t + 1} \right)$. \\
     \STATE Store transition $\left( {s\left( t \right),a\left( t \right),r\left( t \right),s\left( {t + 1} \right)} \right)$ into the experience memory. \\
     \STATE Sample a random mini-batch of $H$ transitions $\left( {s\left( t \right),a\left( t \right),r\left( t \right),s\left( {t + 1} \right)} \right)$ from the experience memory. \\
     \STATE Set $y\left( t \right) = r\left( t \right) + \gamma Q'\left( {s\left( {t + 1} \right),u'\left( {s\left( {t + 1} \right)|{\theta ^{Q'}}} \right)} \right)$. \\
     \STATE Update the parameters $Q\left( {s\left( t \right),a\left( t \right)|{\theta ^Q}} \right)$ in the Q-network $Q\left( {s\left( t \right),a\left( t \right)|{\theta ^Q}} \right)$ by minimizing the loss function ${\bf L}\left( {{\theta ^Q}} \right) = \frac{1}{H}\sum\nolimits_{t \in T} {{{\left( {y\left( t \right) - Q\left( {s\left( t \right),a\left( t \right)|{\theta ^Q}} \right)} \right)}^2}} $.
     \STATE Update the parameters $\mu \left( {s\left( t \right)|{\theta ^\mu }} \right)$ in the deterministic policy network $\mu \left( {s\left( t \right)|{\theta ^\mu }} \right)$ using the policy gradient in Eq.~(\ref{Jtheta}).

     \STATE Update parameters ${\theta ^{Q'}}$ in target Q-network ${\theta ^{Q'}} \leftarrow \tau {\theta ^Q} + \left( {1 - \tau } \right){\theta ^{Q'}}$.
     \STATE Update parameters ${\theta ^{\mu'}}$ in target Q-network ${\theta ^{\mu '}} \leftarrow \tau {\theta ^\mu } + \left( {1 - \tau } \right){\theta ^{\mu '}}$.
     }
     \ENDFOR
     }
     \ENDFOR
 \end{algorithmic}
\end{algorithm}

\subsection{Gradient Descent for Value Function Approximation}\label{section:stochasticgradientdescent}

The goal of RL is to find the relationship between all the states and actions, i.e., the optimal $\theta$ in the DDPG algorithm. We adopt the stochastic gradient descent algorithm to approximate the value function $V\left( s \right)$, which is to minimise the mean-squared error (MSE) between the approximate value function $\widehat V\left( {s,w} \right)$ and the true value function $V\left( s \right)$ defined as follows

\begin{equation}\label{mse}
{\rm MSE}\left( w \right) = {{\mathbb E}_\pi }\left[ {{{\left( {V\left( s \right) - \widehat V\left( {s,w} \right)} \right)}^2}} \right].
\end{equation}

The gradient descent is utilized to find the local minimum which is given by


\begin{equation}\label{stochasticgradientdescent}
\begin{aligned}
\Delta w &=  - \frac{1}{2}\alpha {\nabla _w}{\rm{MSE}}\left( w \right)\\
&= \alpha {{\mathbb E}_\pi }\left[ {\left( {V\left( s \right) - \widehat V\left( {s,w} \right)} \right){\nabla _w}\widehat V\left( {s,w} \right)} \right],
\end{aligned}
\end{equation}
where $\alpha$ is a step-size parameter which determines the learning step size of the gradient descent.

Different from the gradient descent, the stochastic gradient descent algorithm samples the gradient, then represents the value function $V\left( s \right)$ with a linear combination of feature vectors. Hence, we can prove that the stochastic gradient descent algorithm is capable of converging to global minimum.

\begin{theorem}\label{theorem:sgdconvergeglo}
Stochastic gradient descent algorithm for the value function approximation is capable of converging to global optimum.

\begin{IEEEproof}
See Appendix A~.
\end{IEEEproof}
\end{theorem}

\begin{table*}[t!]
\caption{Simulation Parameters}
    \centering
	\begin{tabular}{|c||l||c|}\hline
    	{\bf Parameter}&{\bf Description}&{\bf Value}\\\hline
		$K$ & Antennas at the AP & 4  \\\hline
        $N$ & REs of the RIS & 9 \\\hline
        $2K$ & Number of users & 8  \\\hline
        $B$ & Bandwidth & 20 MHz  \\\hline
        ${{{\rm P}_t}}$ & Transmit power of the AP & 20 dBm  \\\hline
        ${\sigma ^2}$ & Noise power & -138 dBm  \\\hline
        $\alpha$ & Learning rate & 0.01  \\\hline
        $\epsilon$ & Exploration & 0.1  \\\hline
        $\gamma$ & Discount factor & 0.99  \\\hline
        $H$ & Minibatch size & 256  \\\hline
        $U$ & Replay memory & 1000  \\\hline
        $O$ & Hidden layer size & 48  \\\hline
	\end{tabular}
	\label{table2}
\end{table*}

\linespread{1.8}
In Q-learning, the value function approximation evaluates the optimal state-action value function ${Q^ * }\left( {s,a} \right)$, which is given in Eq.~(\ref{mse}). The parameter $w$ is recursively estimated using a stochastic gradient descent. Therefore, $\theta$ is updated as follows

\begin{equation}\label{qlcfa}
\begin{aligned}
{\theta _{i + 1}} &= {\theta _i} - \frac{{{\alpha _i}}}{2}{\nabla _{{\theta _i}}}{\left( {Q_i^ *  - {{\widehat Q}_\theta }\left( {{s_i},{a_i}} \right)} \right)^2}\\
 &= {\theta _i} + {\alpha _i}\left( {{\nabla _{{\theta _i}}}{{\widehat Q}_\theta }\left( {{s_i},{a_i}} \right)} \right)\left( {Q_i^ *  - {{\widehat Q}_\theta }\left( {{s_i},{a_i}} \right)} \right),
\end{aligned}
\end{equation}

\section{Numerical Results}\label{section:numeralresult}

\begin{figure} [t!]
 \centering
 \includegraphics[width=4in]{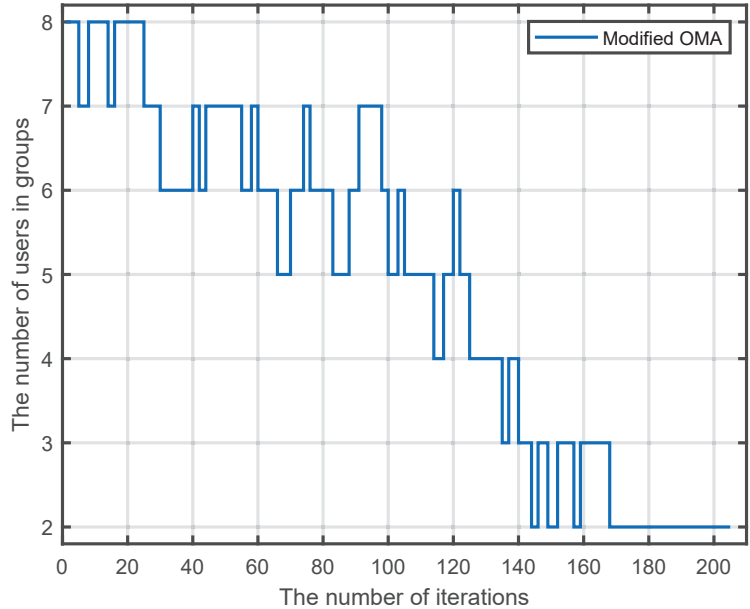}
 \centering
 \caption{The number of users in MOMA clusters different from the optimal partitioning under increased number of iterations (The number of users is 8).}\label{sresult0}
\end{figure}

In this section, we present extensive simulation results to quantify the performance of the proposed MOMA and DDPG algorithms for user clustering and phase shifter design in RIS-aided NOMA downlink networks. We adopt small-scale Rayleigh fading between the AP and users. The simulation parameters settings are as given in Table~\ref{table2} unless otherwise stated. We consider the situation where the positions of the users and the AP are fixed. In our simulations, the positions of users are randomly distributed within a square region with a side length of 500m. The number of users is set to 8, the AP location is in the centre of the square, and the bandwidth is 20MHz. We compare our proposed algorithm with the following three conventional downlink schemes: (1) ``\emph{RIS-aided OMA downlink}" where the RIS is applied for the OMA downlink transmission. (2) ``\emph{Fixed phase shifting RIS-aided NOMA downlink}" where the phase shifts of the RIS are fixed during the learning process. (3) ``\emph{Random phase shifting RIS-aided NOMA downlink}" where the phase shifts of the RIS are random in each time slot during the whole period. All the simulations are performed on a desktop with an Intel Core i7 9700K 3.6 GHz CPU and 16 GB memory. We use the MATLAB$^{\copyright}$ R2019b programming language for the proposed MOMA algorithm and RL algorithms.

\subsection{The convergence of the proposed MOMA algorithm}\label{section:numeralresult1}
Fig.~\ref{sresult0} shows the number of users that are not correctly placed in the optimal user partitioning when increasing the number of iterations during the learning process. According to Fig.~\ref{sresult0}, in the initial learning state, there is a large number of users that are partitioned into wrong clusters. Then, with more iterations, the number of users in the wrong clusters is reduced.

\begin{figure} [t!]
 \centering
 \includegraphics[width=4in]{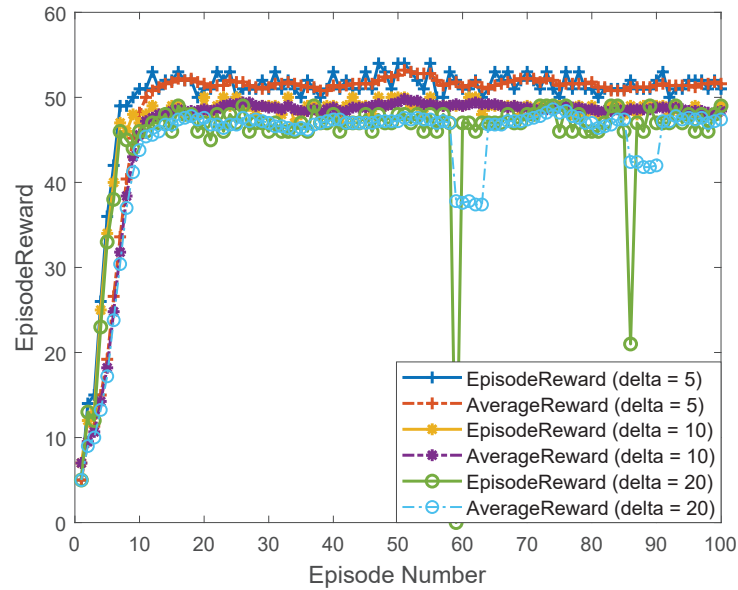}
 \centering
 \caption{The sum reward versus the number of trials for different delta (delta represents the granularity in the action space for changing the phase shifts of the RIS).}\label{sresult1}
\end{figure}

\begin{figure} [t!]
 \centering
 \includegraphics[width=4in]{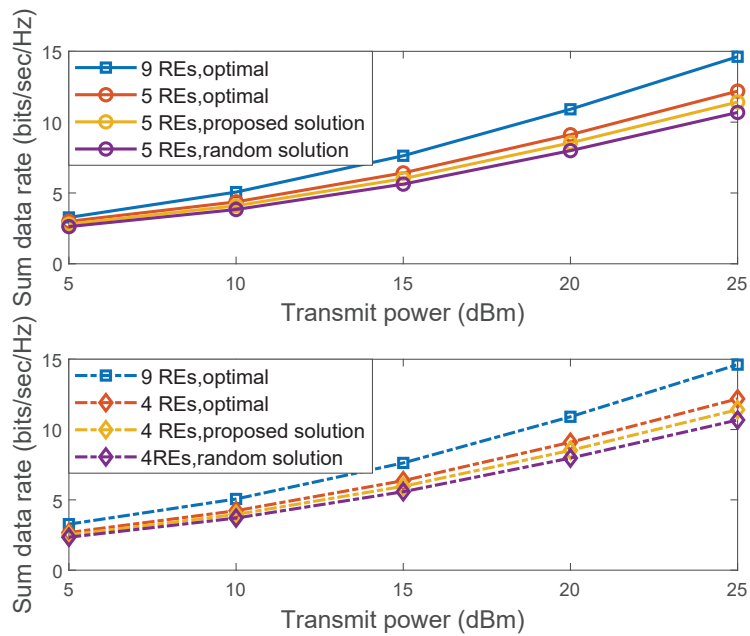}
 \centering
 \caption{The sum data rate versus the AP transmit power.}\label{sresult2}
\end{figure}

\subsection{The performance of the proposed DDPG algorithm}\label{section:numeralresult2}

\begin{figure} [t!]
 \centering
 \includegraphics[width=4in]{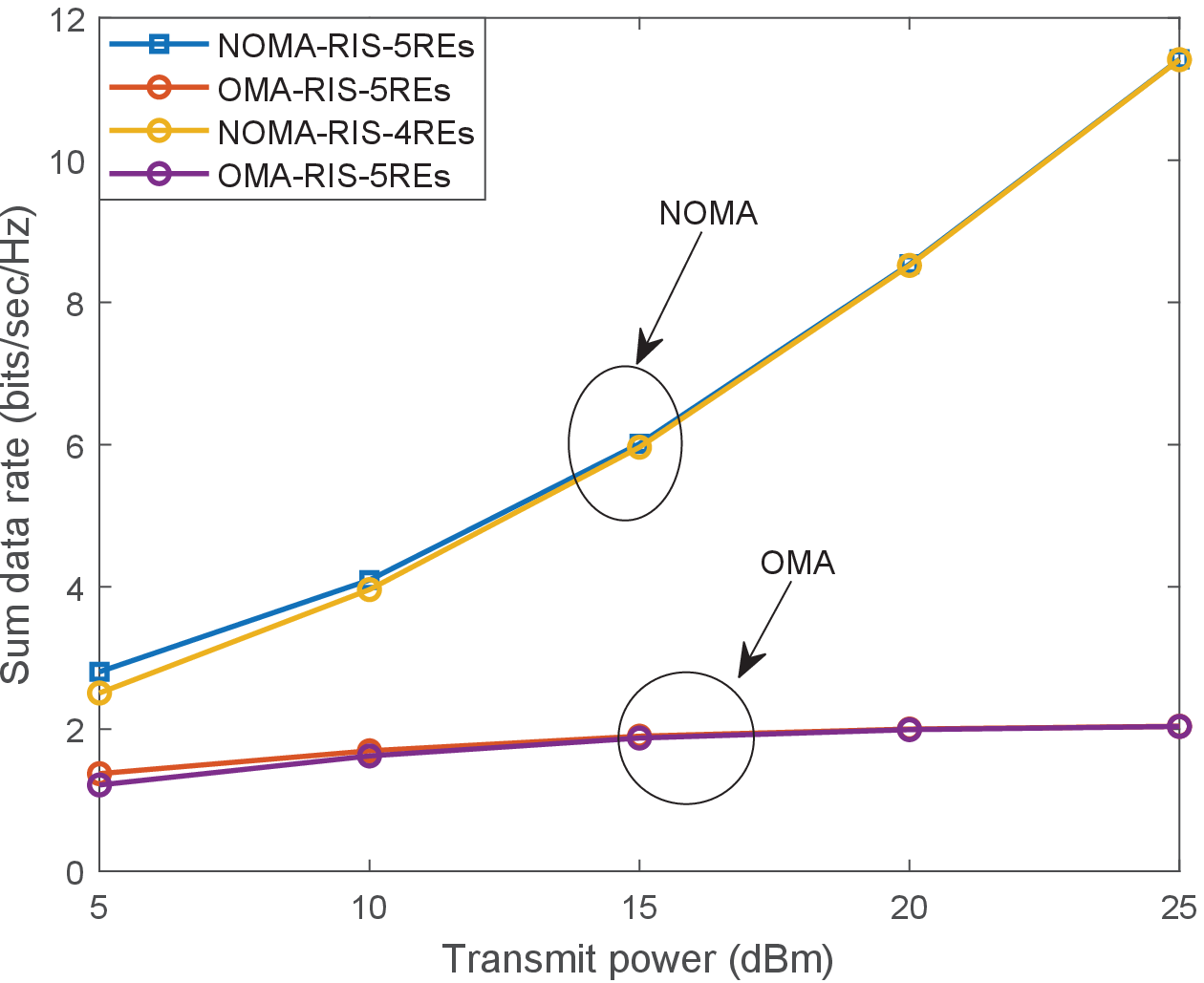}
 \centering
 \caption{The sum data rate versus the AP transmit power.}\label{sresult3}
\end{figure}

\begin{figure} [t!]
 \centering
 \includegraphics[width=4in]{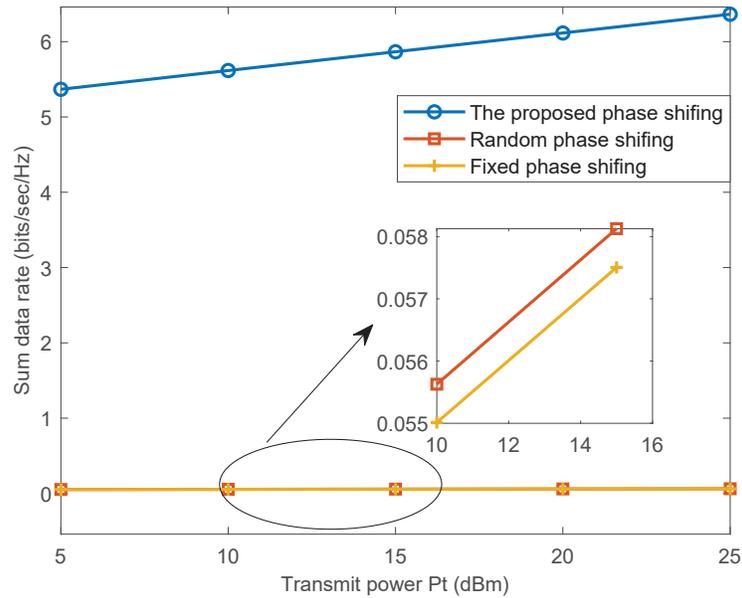}
 \centering
 \caption{The impact of the transmit power on the sum rate.}\label{transmitpower}
\end{figure}

\begin{figure} [t!]
 \centering
 \includegraphics[width=4in]{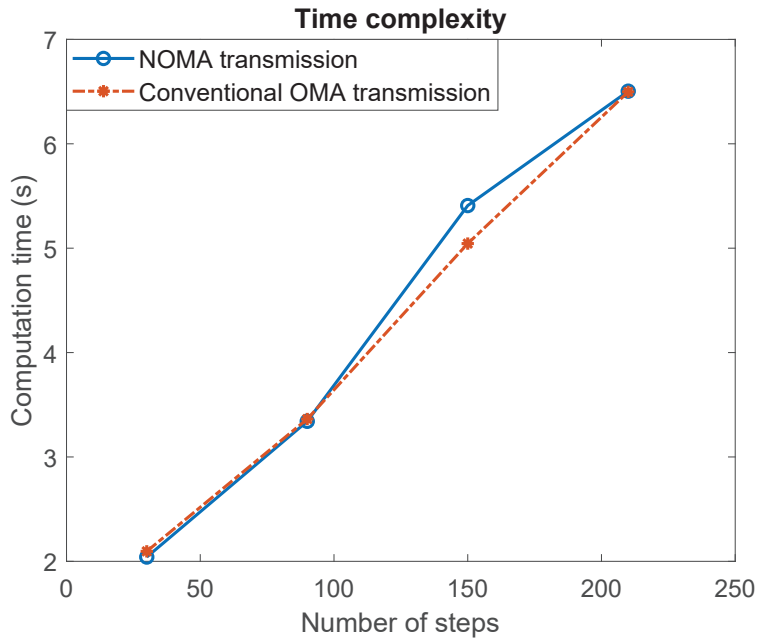}
 \centering
 \caption{The impact of different number of steps on the time complexity.}\label{timecomplexitysteps}
\end{figure}

Fig.~\ref{sresult1} to Fig.~\ref{timecomplexitysteps} presents the performance of the proposed DDPG algorithm for the RIS phase shifter design. Fig.~\ref{sresult1} presents the sum reward versus the number of trials for different delta, where delta is the granularity of the action space for changing the phase shift of the RE. From Fig.~\ref{sresult1}, during the training phase, the sum reward is increasing with the number of trials until it converges to a stable value. This is because the intelligent agent is capable of learning the phase shifts in each trial and remembers the learning history. We can also see that the performance of the proposed algorithm can be improved by reducing the granularity of the phase shifts.

Fig.~\ref{sresult2} shows that the sum data rate grows when increasing the AP transmit power. We calculate the optimal solution by exhaustive search. The optimal solution for the 9 REs case is shown as a benchmark. Since the computation complexity of the proposed DDPG algorithm increases exponentially with the number of REs, we only show the optimal solution for the 9 REs case. Fig.~\ref{sresult2} shows that the proposed algorithm outperforms random phase shifting of the RIS, and achieves performance close to the optimal solution. Meanwhile, when we increase the number of REs in the RIS, the sum data rate grows rapidly, which indicates that increasing the number of REs is an efficient method to improve the performance
of the proposed framework.

Fig.~\ref{sresult3} shows that the proposed RIS-NOMA downlink transmission outperforms the conventional RIS-OMA downlink transmission significantly. As the transmit power increases, the performance gap between NOMA strategy and OMA strategy increases and the sum data rate gap between different numbers of REs reduces.

Fig.~\ref{transmitpower} shows the sum rate versus increased transmit power where the random phase shifting and fixed phase shifting are presented as benchmarks. According to Fig.~\ref{transmitpower}, the proposed phase shifting scheme based on the proposed DDPG algorithm outperforms the random phase shifting and fixed phase shifting schemes. Also from the subwindow in Fig.~\ref{transmitpower}, we observe that the sum rate increases with higher transmit power.

Fig.~\ref{timecomplexitysteps} shows the time complexity (as measured by the execution time in seconds) of the proposed algorithm versus the number of steps. The simulations are performed on a desktop with an Intel Core i7 9700K 3.6 GHz CPU and 16 GB memory and MATLAB$^{\copyright}$ R2019b programming language is used for the proposed algorithm. According to Fig.~\ref{timecomplexitysteps}, the time complexities of both NOMA and conventional OMA increase almost linearly with the number of steps. In addition, we can see from Fig.~\ref{timecomplexitysteps} that the time complexities of NOMA and OMA are very close, which demonstrates that the transmission strategy does not have significant influence on the time complexity.

\section{Conclusion}\label{section:conclusion}

In this paper, a RIS-aided NOMA downlink design framework is proposed. To maximize the sum data rate of the mobile users, a long-term joint optimization problem is formulated, subject to the decoding order of NOMA transmission. Two parameters, namely, user partitioning and RIS phase shifts, are optimized in the maximization problem. For user clustering, a MOMA algorithm is adopted due to its low complexity and fast convergence speed. The users are partitioned into equal-size clusters by the proposed MOMA algorithm. For RIS phase shifting in NOMA downlink network, a DDPG algorithm based phase shifting design scheme was proposed, utilizing the strong fitting ability of neural networks. Both Q-networks and deterministic policy networks are applied in the proposed DDPG algorithm. The effectiveness of the proposed framework and algorithms were illustrated by numerical experiments. Numerical results demonstrate that the performance of the proposed framework can be improved by reducing the granularity of the RIS phase shifts and increasing the number of REs of the RIS.

\section*{Appendix~A: Proof of Theorem~\ref{theorem:sgdconvergeglo}} \label{Appendix:A}
In order to prove {\bf Theorem~\ref{theorem:sgdconvergeglo}}, we formulate the stochastic gradient decent based value function approximation (SGD-VFA) problem as follows

\begin{equation}\label{stgvfa1}
\mathop {\min }\limits_{\omega  \in R} f\left( \omega  \right) = \frac{1}{N}\sum\limits_{i = 1}^N {{\rm MSE}_i\left( \omega  \right)} ,
\end{equation}
where $N$ is the number of total iterations. The formulated SGD-VFA problem is to minimize the total loss function given the approximation parameter in the approximate value function $\widehat V\left( {s,w} \right)$.

To solve the above optimization problem, the SGD algorithm starts with an initial vector ${\omega _0}$ and generates a sequence $\left\{ {{\omega _k}} \right\}$ according to the following equation

\begin{equation}\label{stgvfa2}
{\omega _{k + 1}} = {\omega _k} - \frac{\alpha }{2}\nabla {\rm{MSE}}\left( {{\omega _k}} \right),
\end{equation}
where $\alpha$ is the learning rate. Equation~(\ref{stgvfa2}) can be rewritten as

\begin{equation}\label{stgvfa3}
{\omega _{k + 1}} = \mathop {\arg \min }\limits_{u \in R} \left\{ {{\rm{MSE}}\left( {{\omega _k}} \right) + \left\langle {u - {\omega _k},\nabla {\rm{MSE}}\left( {{\omega _k}} \right)} \right\rangle  + \frac{1}{\alpha }{{\left\| {u - {\omega _k}} \right\|}^2}} \right\}.
\end{equation}

The value function $V(s)$ is linear, therefore, the objective function $f\left( \omega  \right)$ in Eq.~(\ref{stgvfa1}) is also linear. Thus, we have Eq.~(\ref{weightvector3}).

\begin{figure*}[!t]
\normalsize
\begin{equation}\label{weightvector3}
\begin{aligned}
\alpha \left( {f\left( {{\omega _{k + 1}}}  \right) - f\left( {{\omega ^*}} \right)} \right) &= \alpha \left\langle {\nabla {\rm{MSE}}\left( {{\omega _k}} \right),{\omega _{k + 1}} - {\omega ^*}} \right\rangle \\
 &= \left\langle {2\left( {{\omega _k} - {\omega _{k + 1}}} \right),{\omega _{k + 1}} - {\omega ^*}} \right\rangle \\
 &= 2\left( {\left\langle {{\omega _k},{\omega _{k + 1}}} \right\rangle  - \left\langle {{\omega _k},{\omega ^*}} \right\rangle  - \left\langle {{\omega _{k + 1}},{\omega _{k + 1}}} \right\rangle  + \left\langle {{\omega _{k + 1}},{\omega ^*}} \right\rangle } \right)\\
 &= 2\left( {{\omega _k}{\omega _{k + 1}} - {\omega _k}{\omega ^*} - {\omega _{k + 1}}{\omega _{k + 1}} + {\omega _{k + 1}}{\omega ^*}} \right)\\
 &= {\left\| {{\omega _k} - {\omega ^*}} \right\|^2} - {\left\| {{\omega _{k + 1}} - {\omega ^*}} \right\|^2} - {\left\| {{\omega _{k + 1}} - {\omega _k}} \right\|^2}
\end{aligned}
\end{equation}
\hrulefill \vspace*{0pt}
\end{figure*}

To avoid the vanishing of the gradient, the difference between ${f\left( {{\omega _{k + 1}}} \right)}$ and ${f\left( {{\omega ^*}} \right)}$ should be smaller than the gradient, therefore, we have

\begin{equation}\label{weightvector4}
f\left( {{\omega _{k + 1}}} \right) \le f\left( {{\omega ^*}} \right) + \frac{1}{\alpha }\left( {{{\left\| {{\omega _k} - {\omega ^*}} \right\|}^2} - {{\left\| {{\omega _k} - {\omega ^*}} \right\|}^2} - {{\left\| {{\omega _{k + 1}} - {\omega ^*}} \right\|}^2} - {{\left\| {{\omega _{k + 1}} - {\omega _k}} \right\|}^2}} \right).
\end{equation}

From Eq.~(\ref{weightvector4}) and Eq.~(\ref{stgvfa1}) we can obtain Eq.~(\ref{weightvector5}).

\begin{figure*}[!t]
\normalsize
\begin{equation}\label{weightvector5}
\begin{aligned}
{\rm{MSE}}\left( {{\omega _{k + 1}}} \right) &\le {\rm{MSE}}\left( {{\omega ^*}} \right) + \frac{1}{\alpha }\left( {{{\left\| {{\omega _k} - {\omega ^*}} \right\|}^2} - {{\left\| {{\omega _{k + 1}} - {\omega ^*}} \right\|}^2} - {{\left\| {{\omega _{k + 1}} - {\omega _k}} \right\|}^2}} \right)\\
 &\le {\rm{MSE}}\left( {{\omega ^*}} \right) + \frac{1}{\alpha }\left( {{{\left\| {{\omega _k} - {\omega ^*}} \right\|}^2} - {{\left\| {{\omega _{k + 1}} - {\omega ^*}} \right\|}^2}} \right).
\end{aligned}
\end{equation}
\hrulefill \vspace*{0pt}
\end{figure*}

In Eq.~(\ref{weightvector5}), since ${\rm{MSE}}\left( {{\omega _{k + 1}}} \right) \ge {\rm{MSE}}\left( {{\omega ^*}} \right)$, therefore ${\left\| {{\omega _k} - {\omega ^*}} \right\|^2} \ge {\left\| {{\omega _{k + 1}} - {\omega ^*}} \right\|^2}$.

We prove {\bf Theorem~\ref{theorem:sgdconvergeglo}} by contradiction. Consider $\forall k \in \left[ {1,N} \right]$, assuming that the limit of ${\omega _k}$ is $\eta$, i.e., ${\omega _k} \to \eta_k $, thus

\begin{equation}\label{weightvector6}
{\left\| {{\omega _k} - {\omega ^*}} \right\|^2} \ge {\left\| {{\omega _{k + 1}} - {\omega ^*}} \right\|^2} \to {\left\| {\eta_k  - {\omega ^*}} \right\|^2}.
\end{equation}

Assuming that $\eta_k  \notin {W^*}$, where ${W^*} =  \cap _{k = 1}^N{W_k}^*$ is the common space of global minimizers and ${W_k}^*$ is the global minimizer of ${{\rm{MS}}{{\rm{E}}_k}\left( \omega  \right)}$ in Eq.~(\ref{stgvfa1}). $\eta_k  \notin {W^*}$ implies that the limit of ${{\omega _k}}$ belong to ${W_k}^*\backslash {W^*}$, i.e., ${\eta _k} \in {W_k}^*\backslash {W^*}$.

For $u \ne v$, we have ${\eta _u} \in {W_u}^*\backslash {W^*}$ and ${\eta _v} \in {W_v}^*\backslash {W^*}$. Therefore,

\begin{equation}\label{weightvector7}
{\eta _u} \ne {\eta _v}.
\end{equation}

However, during the iterations, if $u = v + 1$ then Eq.~(\ref{weightvector8}).

\begin{figure*}[!t]
\normalsize
\begin{equation}\label{weightvector8}
\begin{aligned}
{\left\| {{\omega _v} - {\eta _u}} \right\|^2} \ge {\left\| {{\omega _u} - {\eta _u}} \right\|^2} \ge {\left\| {{\omega _{u + 1}} - {\eta _u}} \right\|^2} \ge  \cdots  \ge {\left\| {{\omega _{u + m}} - {\eta _u}} \right\|^2}\mathop  \to \limits^{m \to \infty } 0.
\end{aligned}
\end{equation}
\hrulefill \vspace*{0pt}
\end{figure*}

According to Eq.~(\ref{weightvector8}), we obtain ${\omega _v} \to {\eta _u}$, therefore

\begin{equation}\label{weightvector9}
{\eta _u} = {\eta _v}
\end{equation}

Clearly, Eq.~(\ref{weightvector7}) and Eq.~(\ref{weightvector9}) contradict each other. Thus, the assumption that $\eta_k  \notin {W^*}$ is not true. Hence, the sequence $\left\{ {{\omega _k}} \right\}$ converges to a global minimizer.

The proof is complete.

\bibliographystyle{IEEEtran}
\bibliography{mybib}

\end{document}